\documentclass[aps,amsmath,prb,a4paper,floatfix,twocolumn]{revtex4}
\usepackage{graphicx}
\newcommand{\met}{\frac{1}{2}}
\newcommand{\beq}{\begin{equation}}
\newcommand{\eneq}{\end{equation}}
\newcommand{\bea}{\begin{eqnarray}}
\newcommand{\enea}{\end{eqnarray}}
\newcommand{\bc}{\begin{center}}
\newcommand{\ec}{\end{center}}

\newcommand{\lvt}{ L_\varphi}
\newcommand{\tif}{ \tau _\varphi}
\begin{document}

\title{Mesoscopic conductance fluctuations in YBa$_2$Cu$_3$O$_{7-\delta}$ grain boundary
Junction at low temperature}

\author{A. Tagliacozzo$^1$, F. Tafuri$^{2,3}$, E. Gambale$^{1,2}$, B.Jouault$^4$, D.Born$^{2,3}$, P.Lucignano$^{1,5}$, D. Stornaiuolo$^1$,F.Lombardi$^6$, A. Barone$^1$, B.L. Altshuler$^7$}

\affiliation{$^{1}$ Coherentia INFM-CNR  and Dipartimento di Scienze Fisiche, Universit\`{a} di Napoli Federico II, Italy } 
\affiliation{$^{2}$ Dip. Ingegneria dell'Informazione, Seconda Universit\`{a} di Napoli, Aversa (CE), Italy}
\affiliation{$^{3}$ NEST CNR-INFM  and Scuola Normale Superiore, I-56126 Pisa,Italy }
\affiliation{$^{4}$ Universit\'e Montpellier 2, Groupe d'\'Etude des
 Semiconducteurs,  and 
 CNRS, UMR 5650, cc074,  pl. Eug\`ene Bataillon, 34095 Montpellier cedex 5, France}
\affiliation{$^{5}$ SISSA  Via Beirut 2-4, 34014 Trieste, Italy }
\affiliation{$^{6}$ 
Dept. Microelectronics and Nanoscience, MINA, Chalmers University of Technology, 41296 Gšteborg, Sweden }
\affiliation{$^{7}$ Physics Dept. Columbia University, New York NY 10027 and  NEC Laboratories America INC, 4 Independence Day, Princeton, NJ 08554,USA}

\date{\today}

\begin{abstract}
The magneto-conductance in YBCO  grain boundary  Josephson junctions,
 displays fluctuations at  low  temperatures of mesoscopic origin.
The morphology  of the junction  suggests that 
transport  occurs in narrow channels across  the  grain boundary line, 
with  a large  Thouless energy. Nevertheless
the measured fluctuation amplitude  decreases quite slowly when  increasing 
the  voltage up to values  about  twenty times the Thouless 
energy, of the order of the nominal superconducting gap.
  Our  findings  show the coexistence  of 
supercurrent and quasiparticle current in the junction conduction even at high 
nonequilibrium conditions. Model calculations confirm the reduced role 
of quasiparticle relaxation at temperatures up to 3 Kelvin.
\end{abstract}

\maketitle

\section{Introduction}

In high critical Temperature Superconductor (HTS) junctions, including grain boundary (GB) structures,
it is well established that various interplaying mechanisms contribute to transport with 
different weights still to be completely defined in a general and consistent framework 
\cite{kirtley,hans,rop}.  
As in all non-homogeneous systems, the barrier region will significantly contribute to determine the transport 
properties across the structure. What is peculiar of HTS is the complicate material science entering in the 
formation of the physical barrier microstructure. This  will depend on the
type of device and the fabrication procedure. The material science complexity of HTS 
may also result in different precipitates and inclusions present at interfaces and grain boundaries, 
and in some type of inherent lack of uniformity of the barriers \cite{hans}.
All this has turned into some uncertainty about the nature of the barrier and has originated
various hypotheses on the transport properties. The most wide-spread models are basically all in-between two extreme ideas 
\cite{gross,gross1,gross2,gross3,buhrman,buhrman1,hans}:
on the one hand resonant tunnelling through some kind of dielectric barrier \cite{gross,gross1,gross2,gross3}, on the other, 
especially in GB junctions, a barrier composed of thick insulating regions separated by conducting channels, 
which act as shorts or microbridges \cite{buhrman,buhrman1}.
In most cases the interface can be modeled as an intermediate situation between the two limits mentioned above. A transition from one extreme to the other can therefore take place. What is
unfortunately missing is a way to describe this tuning-transition through reliable and well defined barrier 
parameters (for instance the barrier transparency). 
The predominant d-wave order parameter symmetry (OPS) is another important  \cite{kirtley}, 
to which  a large part of the phenomenology has been clearly associated \cite{kirtley,hans,rop,wendin,wendin1}. D-wave OPS implies  the presence of antinodal (high energy), and nodal (low energy) quasi-particles  in the conduction across junctions and  the absence of sharp gap features  in the density of states of the weak link.
Recently, low temperatures measurements have proved macroscopic quantum tunneling (MQT) in YBCO GB junctions, stimulating  
novel research on coherence and dissipation in such complex systems \cite{nuovo,mqt,mqt1}.

In this work we report on an investigation of magnetoconductance at low temperatures for the same 
type of biepitaxial GB junctions \cite{tem}, used for the MQT experiments. These structures are very flexible and versatile,
guaranteeing on the one hand low dissipation\cite{mqt,mqt1} and on the other a reliable way to
 pass from tunnel-like to diffusive
transport on the same chip by changing the interface orientation\cite{nuovo,tem}.
We give direct evidence of the role played by narrow conduction channels across the GB. These channels may 
have different sizes and distributions and obviously a different impact on the transport properties. 
 When increasing 
applied voltage, mesoscopic  conductance fluctuations\cite{aronov,lee,lee2,fukuyama,today} appear in our samples,
at low temperatures, not dissimilar from what is usually observed in normal 
narrow metal samples \cite{pierre}. 
We expect  that, in our sample, typical sizes of the current-carrying constrictions $L_x$ and $L_y$ range from 50 nm to 
100 nm  and, as a consequence, Thouless energy $E_c$ ( see Table \ref{tab1} ) turns out to be quite large, when compared to the values usually measured 
in traditional normal metal artificial  systems \cite{oudenaarden}. The 
mesoscopic effects persist at voltages about twenty times larger than the Thouless energy.
"Novel"  mesoscopic issues that  emerge from the analyses carried out
in the present work are tightly connected to the  nature of  the GB systems: 
a) a smooth crossover appears to exist from the coherent conduction mostly driven by the supercurrent, to the magnetoconductance driven by quantum coherent diffusion 
of quasiparticles across the mesoscopic area, when the voltage at the junction increases; 
b) in analogy to  pairs, quasiparticles also appear to have a large  phase coherence as 
proved by the shape of the power spectrum of the conductance fluctuations, 
up to temperatures of 3K; 
c) the voltage drop appears to be concentrated at 
the GB, and non equilibrium  does not  affect  substantially the
mesoscopic interference over a wide area  about $1\mu m^2$.

This work builds upon a previous report, where the main ideas have been illustrated \cite{tagliacozzo}. 
Herein a more complete  analysis of the experimental data is  carried out.  We have applied the "protocol" established in the last 20 years on semiconducting and
normal metal nanostructures to our system and we have extracted the characteristic lengths and scaling energies.

In Section II we give some details about the sample fabrication. 
By  presenting the magnetic mesoscopic  
fingerprints of our sample  in  Section IIIA, we collect evidences of the mesoscopic character of the conductance 
fluctuations that we have measured. In section IIIB we derive from  the ensamble average of the 
fluctuations  the variance of the conductance, which is presented in section IIIC.

In Section IV we show the conductance 
autocorrelation for different  magnetic fields in an  intermediate  voltage range.
By analyzing the power spectral density we estimate the phase coherence length $L_\varphi $ which is found to be  $ \lesssim L$, at intermediate voltages  ( $V = 14\div 18 $ mV).
 The conductance autocorrelation at different voltages allows us to interpret the role of nonequilibrium by defining  the voltage dependence of  the phase coherence length, as discussed in Section V.
The discussion of the results can be found in Section VI. 
Our simple model theory well  accounts for the  experimental results and clarifies the survival of non locality in the quantum diffusion in 
presence of a large voltage bias. 
Table \ref{tab1} gives additional information on the planning of this work. 
Our conclusions can be found in Section VII.

\begin{widetext}
\begin{table}[htdp]
\caption{Summary of the results:}
\begin{center}
\begin{tabular}{|c|c|c|c|}
\hline \hline Topic & Relative measurement  &  Extracted parameters &  Estimates  \\ \hline\hline
Junction geometry            &  Magnetic pattern:  $I_C \:  vs.\: H $ in Fig.(\ref{figp1}) & $L_x\sim L_y\sim50nm$ & $E_c\sim1mV$           \\
\hline
Transport properties        & $R_N $ in  Fig.(\ref{1res}) and $I_c R_N$  in Fig.(\ref{daniela}) & $R_N=400 \Omega$,  $T  \gtrsim 257 mK$ & $I_c R_N\sim E_c/e $\\
\hline
Mesoscopic Fingerprints & Resistance fluctuations in Fig.s(\ref{fig3c},\ref{figpio})    & $var[g] \lesssim 1$ &                             \\
\hline
Phase coherence length & Autocorrelation vs $\Delta H$ in Fig.(\ref{fig7}), PSD in Fig.s(\ref{fpsd1},\ref{emilia2}) & $L_\varphi \lesssim 1 \mu m$ & $L_\varphi \sim V^{-1/4}$ \\
\hline
Coherent phase breaking time        & Autocorrelation vs $\Delta V$ in Fig.(\ref{autocorri})                                 & $\tau_\varphi \sim 400\: ps$& $D \tau_\varphi \sim L_\varphi^2$    \\
\hline
\end{tabular}
\end{center}
\label{tab1}
\end{table}%
\end{widetext}

\begin{figure}[htb]	
\begin{center}		
\includegraphics[width=6cm,angle=270]{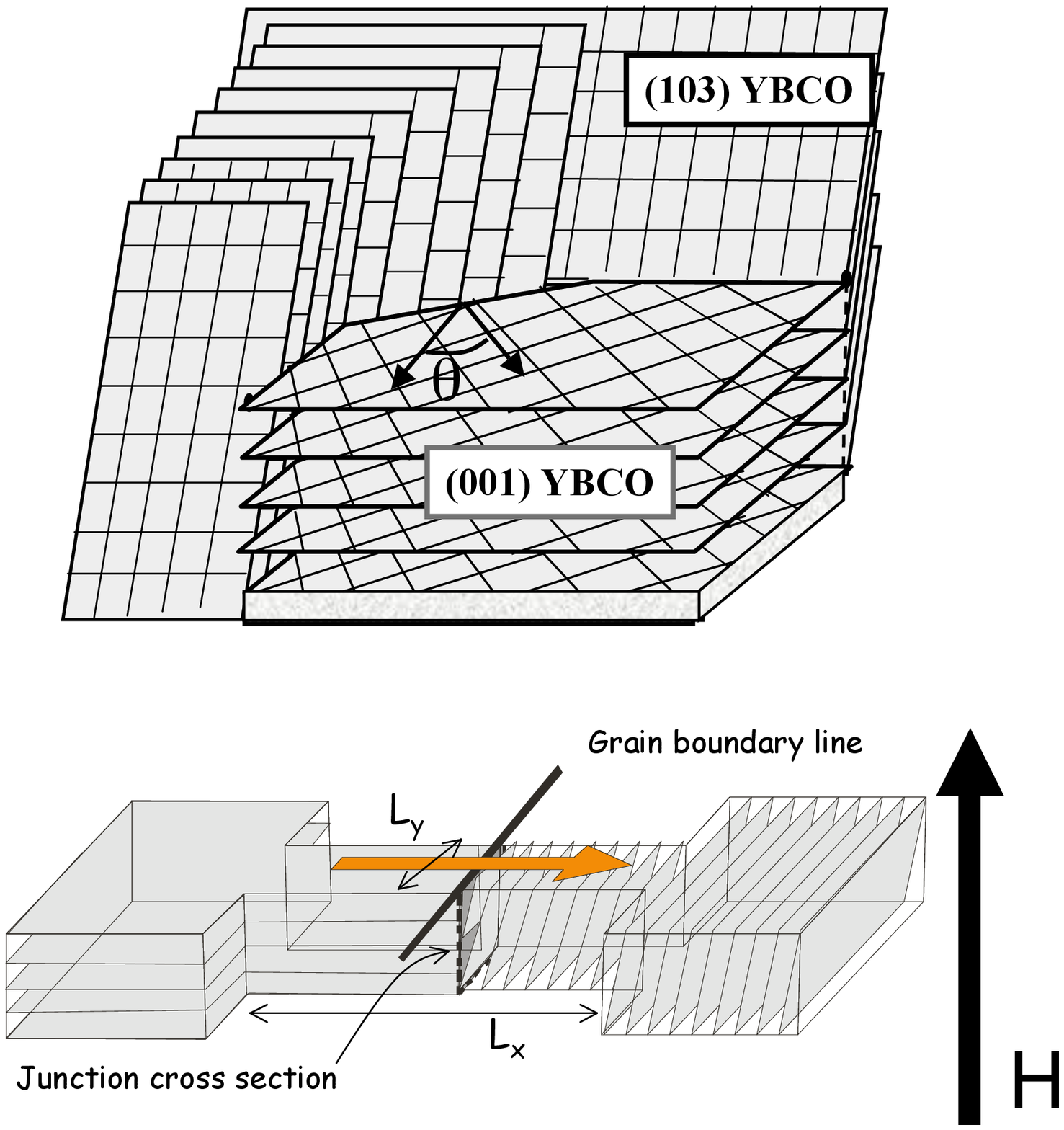}	
\end{center}	\caption{a) Sketch of the YBCO  biepitaxial GB junction used in this experiment. b)Geometry of the model system  of the current-carrying  constriction.}
	\label{sketch}
\end{figure}
\begin{figure}[htb]	
\begin{center}		
\includegraphics[width=6cm]{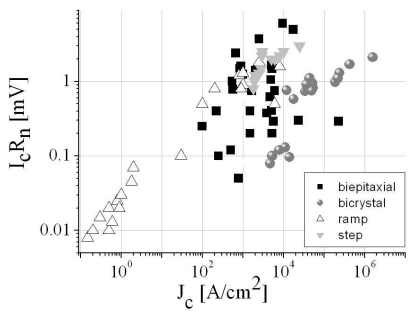}	
\end{center}	\caption{ ({\sl color online}) The product  $I_c R_N$  is reported 
as a function of the critical current density $J_c$. Data are collected from Ref.s \cite{hans,rop,dimos,hilgenAPL,ivanov,char} }
	\label{daniela}
\end{figure}

\section{fabrication and average transport properties of the sample}

The $GB$ Josephson junctions are obtained at the interface between a  (103)YBa$_{2}$Cu$_{3}$ O$_{7-\delta }$ ($YBCO$) film grown on a (110) SrTiO$_3$
substrate and a $c$-axis film deposited on a (110) CeO$_2$  seed layer (see Fig.(\ref{sketch})). The presence of the CeO$_2$ produces an additional 45$^\circ$
in-plane rotation of the $YBCO$ axes with respect to the in-plane directions of the substrate \cite{nuovo}. 
The  angle $\theta$ of the grain boundary relative to the substrate $ a, b $ axes is defined by suitably 
patterning lithographically the CeO$_2$ seed layer (see Fig. (\ref{sketch} a)). Details about the fabrication process and a wide characterization of superconducting properties can be found
elsewhere \cite{tem,nuovo}.  The interface orientation 
can be tuned to some appropriate transport regime, evaluated  through the normal state resistance $R_N$ and critical current density (J$_C$).  
Typical values are reported in Fig.(\ref{daniela}) and compared with data available in literature \cite{rop}.

\begin{figure}[htb]	
\begin{center}		
\includegraphics[width=6cm]{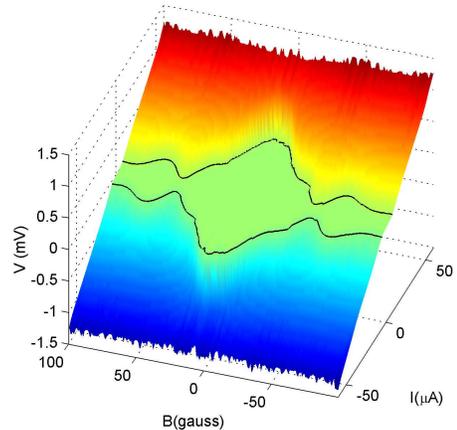}	
\end{center}	\caption{({\sl color online}) The I-V characteristics as a function of  magnetic field.}
	\label{figp1}
\end{figure}

In the tilt cases J$_C$ $\approx$  10$^3$ A/cm$^2$ and $\sigma$$_N$=$1/(R_N A)$ $\approx$ 
0.2(m$\Omega$cm$^2$)$^{-1}$, both measured at T = 4.2 K (A is the junction cross section).
Twist  GBs junctions are typically characterized by higher values of J$_C$
in the range 0.1-4.0 x 10$^5$ A/cm$^2$ and
 $\sigma_N \approx$10 (m$\Omega$cm$^2$)$^{-1}$ (at T = 4.2 K).

We have selected YBCO grain boundary junctions and measured their 
I-V curves  at low temperatures T,  as a function of the magnetic field H
applied  in the direction orthogonal to the plane of the junction. 
In HTS junctions, the correlation between the
magnetic pattern and the current distribution profile along the junction
is made more complicate by the d-wave order parameter symmetry (OPS), which
generates anomalous magnetic response especially for faceted interfaces\cite{kirtley,hans,rop,mannhartprl}.
Additional deviations are expected because of the presence of the second
harmonic in the current-phase relation \cite{wendin,wendin1}. In Fig.(\ref{figp1}) we report the
magnetic field dependence of the IV characteristics of the maximum Josephson current of the  of the junction that we have extensively
investigated in this work (with barrier orientation $\theta$=60$^\circ$). 
This angle gives the maximum $J_C$\cite{nuovo}. 
The magnetic response presents a maximum of the critical current at zero field and two almost symmetric lobes for negative and positive magnetic fields respectively. At higher magnetic fields (above 100 G or below -100 G), the critical current is negligible. The flux  periodicity  is  roughly  consistent with the size  we expect for our microbridge (50-100 nm), since the London penetration depth in the off-axis electrode is larger than the one in c-axis YBCO films, of the order of microns (see Ref. \cite{mqt1} and \cite{tafuriprb00} for instance). Experimental data can be compared with the ideal Fraunhofer case  in the crudest approximation without taking into account the presence of a second harmonic or any specific feature of HTS. Even if deviations from the ideal  Fraunhofer pattern are present, they can be considered to some extent minor if compared  with most of the data on HTS grain boundary Josephson junctions, which present radical differences. We can infer an uniformity of the junction properties approximately on an average scale of 20-30 nm, which is remarkable if compared with most results available in literature. Even if we cannot draw any conclusion on the current distribution on lower length scales, we can rule out the presence of  impurities of large size along the width of the active microbridge. In fact, were there more than  one active microbridge, the  current  of  each of them would  add  in  parallel  and  the  pattern  would   present  other periodicities  referring  to  the  area  enclosed  between  the conduction channels (Ref.\cite{barone}).
The I-V characteristics of the HTS Josephson junctions still present features, 
which cannot be completely understood in terms of the classical approaches used to describe 
the low critical temperature superconductors Josephson junctions. 
These are frequently observed and often referred in literature as unconventional features \cite{kirtley,hans,rop}.
Examples are\cite{rop}: a) I$_C$R$_N$ values are much lower than the  gap value $\Delta$;  b) the shape 
of the I-V strongly depends on the critical current density; c) I-V curves show significant deviations 
from the Resistively Shunted Junction model (RSJ); d) there is a poor consistency between 
the amplitude of the hysteresis and the extracted values of the capacitance, 
when compared to low-$T_C$ superconductor junctions.

\begin{figure}[htb]	
\begin{center}		
\includegraphics[width=6cm]{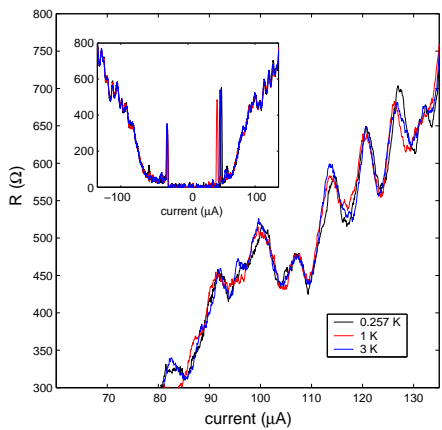}	
\end{center}	\caption{ ({\sl color online})  Main: zooming on the  oscillations of  $R \:vs.\: I $ above $I\approx  80 \mu A$. Inset: Resistance as a function of applied bias current for three different temperatures: $257 mK, 1K, 3K$.}
	\label{1res}
\end{figure}

At low temperatures,   the resistance 
vs applied current  $R(I)$, as derived from the I-V characteristics is rather 
temperature insensitive, while  the critical current $I_C$ maintains a sizable temperature dependence. 
In  the inset of Fig.(\ref{1res}) we show the resistance  $R(I)$  at 
zero magnetic field for three temperatures: $257mK$, $1K$ and $3K$. 
$R(I)$ obviously 
vanishes in the Josephson branch and displays a
sharp peak  when   switching  to/from  the finite voltage conductance.
The data are displayed in order to show the critical current at $I>0$ and 
the retrapping current at $I<0$.  
In Fig.(\ref{1res},main-panel) we show a blow up of the resistance  $R(V)$
in a range of  voltage values
$V$ between $ 0.25 \: mV $  and $ 30 \: mV$ at zero magnetic field and 
for different  temperatures  $T=257 mK, 1K,3K $. 

Measurements have been taken after  different cool-downs in the time lapse of  
two years  to study the sample dependent  properties. 

The average 
resistance  in the range of voltages $
V\approx 10 mV \div 15 mV $  has been stable for about eighteen months 
at  $ \sim 180 \:\Omega $ and has increased in the last year 
up to about $430 \: \Omega $.
These changes should be attributed to aging of the diffusion properties
at the grain boundary. However, in the meantime, no significant change of the $I_C$ has been detected.
Only one sample was available with such a reduced width. The steady progress in nanotechnology
will probably lead to the realization of reliable microbridges of nominal width of a few hundred nanometers,
from which it will be easier to have junctions with transport carried by very few mesoscopic channels.
We finally signal a strong similarity of the I-V and dI/dV-V curves of our junctions with those 
from sub-micron YBaCuO junctions, reported in Ref.\onlinecite{herbstritt}. 

\section{ Magnetoresistance and mesoscopic fingerprints}
\subsection{ Resistance fluctuations}

According to what reported in the previous Section, we figure out that most of the current in the junction substantially flows across a single nanobridge of characteristic size $L_x <~$ 100nm. Possible lack of spatial uniformity of the current distribution on a scale of less than 20 nm does not affect the arguments developed below.

We call $x$ the flow direction and $y$ the direction perpendicular to  the nanobridge, as shown in Fig.(\ref{sketch}b). 
The diffusion coefficient  $D$  in $YBCO$ is expected to be
 $\sim 20\div  24 \: cm^2/sec $\cite{gedik}. By estimating the 
Fermi velocity $ v_F \approx 7 \times 10^{7} \: cm/sec $ for optimally 
doped $YBCO$, we conclude that the mean free path $\ell$ is smaller than the 
size of the nanobridge.
Therefore we argue that the  transport in the junction is diffusive, which is confirmed by the observation of the  resistance fluctuations.

The Thouless energy, as derived from the expected size of the nanoconstriction, is  $E_c  = \hbar D / L^2  \gtrsim 1\: meV $. This value is confirmed by
our measurements as discussed in Section III.  
The  number of transverse  scattering channels 
in the constriction for a fixed cross section $ A $ is approximately ${\cal {N}}_{ch} \approx k_F^2 A \sim  5 \cdot10^4 $.  $A\sim 100\times 100 \: nm^2 $ is given by the product of the thickness of the film and the width of the channel. 
Hence, quantization of transverse  levels in the bridge does not seem to play any role even at the lowest temperatures investigated.
 Indeed,  $k_B T >> \delta   \sim  {E_c} /{\cal{N}}_{ch} \approx  0.1 \mu eV $, where $\delta $ is 
the  mean  energy level  spacing.
As a consequence,   our system  can be thought  as a   disordered bridge  in   the diffusive limit $\ell <L_x$.

\begin{figure}[htb]	
\begin{center}		
\includegraphics[width=6cm]{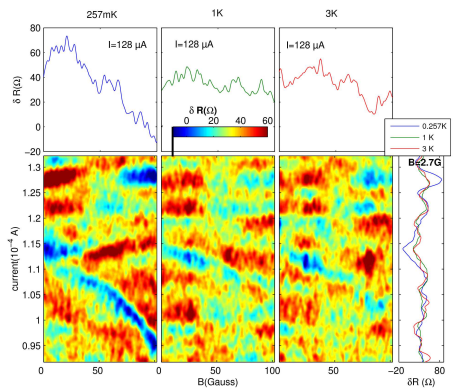}	
\end{center}	
\caption{({\sl color online}) Color plot of the resistance fluctuations  $\delta R(H,I)$ as a function of the applied current  and of the magnetic field for three temperatures: $257 mK$, $1K$, $3K$. The right panel shows $\delta R(H=2.7G,I) \: vs. \: I $ for the three different temperatures. The three top panels show  single magnetoconductance trace for each temperature at fixed bias current (indicated in the labels).}
	\label{fig3c}
\end{figure}

We concentrate on the marked non periodic fluctuations of the  resistance at finite  voltages, with magnetic field in the range  $ B = -100 \div 100 G $.
An example of the  magnetoresistance fluctuations is reported  in Fig.(\ref{fig3c}) top panel for $T=257mK, 1K, 3K $.

The fluctuations are  not related to the magnetic dependence of the  Josephson critical current $I_c (H)$ at voltages $V >2mV $.

In order to avoid  trapping  of flux which may occur when  increasing the temperature,  especially at the higher fields, the   data shown here refer to a single  cool-down.
Occasionally  the pattern has still  a slight  deviation from reproducibility within one single cooling bath, which  could be due to finite relaxation in the
spin orientation of paramagnetic impurities.
CONFERMARE
The resistance pattern derived from our four terminal measurement does not  show any  mirror symmetry $R_N(H) \neq R_N(-H)$. 
Below 1K   there is little temperature dependence.
The amplitude of the fluctuations decreases between $T=1 K$ and $T= 3 K $. 
They are sample dependent, as different cool-downs provide  different patterns. All these features, as well as the ones described below, strengthen the conclusion that they are mesoscopic  fluctuations.  

The  color plot  of the resistance fluctuations  in the $I,H$ plane 
provides  the fingerprints of our sample. 
The deviation from the  average  $\delta R= R-\overline R $ 
is shown in  Fig.(\ref{fig3c}) (three color-plot panels) as a color plot ({\sl color online}),
for three different temperatures, $ 257mK,  1 K,  3K$.  $\overline R$ is the average 
resistance performed over the full  range of magnetic fields.
The pattern  keeps its shape within one single cool down and the  contrast of the colors increases in lowering the temperature.
The color scale is such that the dark red color refers to  resistances significantly larger than the average, while the dark blue color  refers to resistances significantly smaller then the average. 
The data have been filtered by gaussian convolution to get rid  of the  underlying white noise. 

Despite the small equilibrium   thermal  length $L_T  = \sqrt { \hbar D / k_B T }  \sim 0.14 \: \mu m $ at $ 1\: K $, 
there is a clear persistence of the fingerprints  up to $T=3K$. This  suggests that  the strong non-equilibrium conditions induced by the applied voltage do not allow  the thermalization of the carriers in the sample. Indeed, our results    do not change qualitatively  up to  $ 1.5 \: K $ , and transport can be classified as  non equilibrium  quantum diffusive because  $    L_\varphi \lesssim L_x $.
Here  $L_\varphi $  is  the phase  coherence  length for  carriers  diffusing  in   the junction area and is $ \lesssim 1 \mu m $ (see Section IV).

\subsection{Ensemble average}

 The variance $var[g]$  is the 
ensemble average of the  amplitude squared of the conductance fluctuations, 
$ \langle (\delta  g )^2 \rangle$, where $ g = G /(2e^2/h )$ is the 
dimensionless conductance. 
We analyze the fluctuations of the conductance  obtained by  averaging over runs at different magnetic fields up to 
$100 G $, at different voltages.  Here we argue that  this average  can be taken as an acceptable 
ensemble average  and provides a {\sl bona fide} information about the   variance $var[g]$   of the conductance and  its autocorrelation.   To justify our statement, we have to show that  
the Cooperon contribution to the variance is not significantly  influenced by $H$ up to at least  $100 \: G $.
The variance  of the conductance at equilibrium at temperature $T$ can be  calculated as \cite{stone} 
\begin{widetext}
\beq
var[g(H,T) ] =\left . \frac{4s^2}{\pi ^4} \int \frac{d\Delta E }{2k_B T } \:
f\left [ \frac{\Delta E }{2k_B T} \right ] \: \left  [ F_D (\Delta E,H_1,H_2 ) + F_C  (\Delta E ,  H_1,H_2 )\right ] \right |_{H_1=H_2=H}
\label{var} 
\eneq
where $s$ is the spin degeneracy, $f(x) = (x\: coth x -1)/sinh ^2x $ and $F_{D,C}$  are the Diffuson and Cooperon autocorrelation functions, respectively. They can be rewritten  in terms of the eigenvalues $\lambda _\alpha^{D,C}$ of  the diffusion equation\cite{rammer}: 
\beq
\left [ -D\left (-i\vec{\nabla} +\frac{e}{\hbar c}
\left( \vec{A}_1\pm \vec{A}_2 \right)\right)^2 
 +\frac{1}{\tau_{in}}  -i \frac{\Delta E}{\hbar} \right ]\:
\psi _\alpha = \frac{\lambda _\alpha^{D,C}}{\tau} \:
\psi _\alpha \:\: .
\label{eqc}
\eneq 
Here $D\tau = l^2/d  \sim 4 \cdot 10^{-12} cm^2 $ (where $d$ is the effective dimensionality ) and  $\tau _{in} $ is the inelastic relaxation  time  
($ \tau _{in} >> \tau   \sim 0.2 \:  psec $). $A_1,A_2$ are the vector potentials ($H_1$ and $H_2$ the magnetic fields) influencing the outer and inner conductance  loops respectively and the $+/-$ sign refers to the $C/D$ propagator respectively.   We have:  
\bea
F_{D,C} = L^{-4} (D\tau )^2 \:
\sum _\alpha \left [ \frac{1}{\left |\lambda _\alpha^{D,C} \right |^2 } 
+
\met \Re e  \frac{1}{{\lambda _\alpha^{D,C}} ^2 } \right ] \:\: .
\enea
At zero temperature only $\Delta E = 0 $ contributes to the integral in Eq.(\ref{var}), so  that the eigenvalues become real. In  the  evaluation of the variance,  $H_1=H_2=H$ implies that  the Diffuson eigenvalues  become  insensitive of   the magnetic field.  Instead, the Cooperon  eigenvalues depend on $2H $ and can be written in analogy with the Landau levels energies. It follows that :
\beq
F_C (\Delta E = 0 , H) \sim 
 \frac{3}{2 } \: \sum _{n=0}^{n_{max}}    \frac{1}{
\left (n+\met + \frac{1}{\omega _H \tau _{in} }\right )^2}
\sim \frac{3\pi ^2 }{4} \:
\left (1 -  \frac{H}{H_o }\right )  \:\: ,
\label{vcin}
\eneq   
\end{widetext}
where  $ \omega _H = 4eHD/\hbar c $.
Eq.(\ref{vcin})  defines a decay threshold field  of  the Cooperon   $H_o \sim  \pi h c / ( 12 eD \tau  ) $, which  derives  from the truncation of the  sum  over the orbital quantum number $n$  at   $n_{max} \approx  \hbar /m\omega _c l^2  $ ($\omega _c = eH/mc $ is the cyclotron frequency). 
This limitation  is required by    quantum  diffusion  ($ \langle r^2 \rangle _{n_{max}} > l^2 $). The large value of  $ n_{max} $, In our case ($ \sim 10^2 $ for $H < 600 \: G $), determines $H_o  \gtrsim 2.5 \: Tesla $ which is far beyond the field strengths  that can be  applied to our sample without trapping  flux due to vortices. 
This confirms that averaging over the interval of $H $  values $H\in (-100 \div 100\:  G )$ is equivalent to a sample average, without  introducing  significant  field dependencies. 
In the following the ensamble average will be denoted by the symbol $\langle ... \rangle_H$.
As it is shown in the next Section, the typical magnetic field scale that  arises from the autocorrelation of the conductance is $\sim 10 \: G $, much 
smaller than the interval over which the average is performed.

In the rest of the paper, we will generically  denote  the  conductance autocorrelation, which is an  extension of  Eq.(\ref{var}),   by $K_g $.  This quantity depends on many variables: $ T, H\equiv (H_1 +H_2)/2, \Delta H \equiv H_1-H_2, V \equiv (V_1 +V_2)/2, \Delta V \equiv  V_1-V_2 $. When no ambiguity arises,  we have taken the liberty  to list  just the parameters relevant  to the ongoing discussion,  in order  to simplify  the notation. 

\subsection{ Variance of the conductance  and different  voltage regimes}
The conductance  is derived from the $I/V$ characteristic.  We have  checked the behavior of the  differential resistance, measured through  a standard lock-in method, and we have found qualitatively similar results.  

\begin{figure}[htb]	
\begin{center}		
\includegraphics[width=\linewidth]{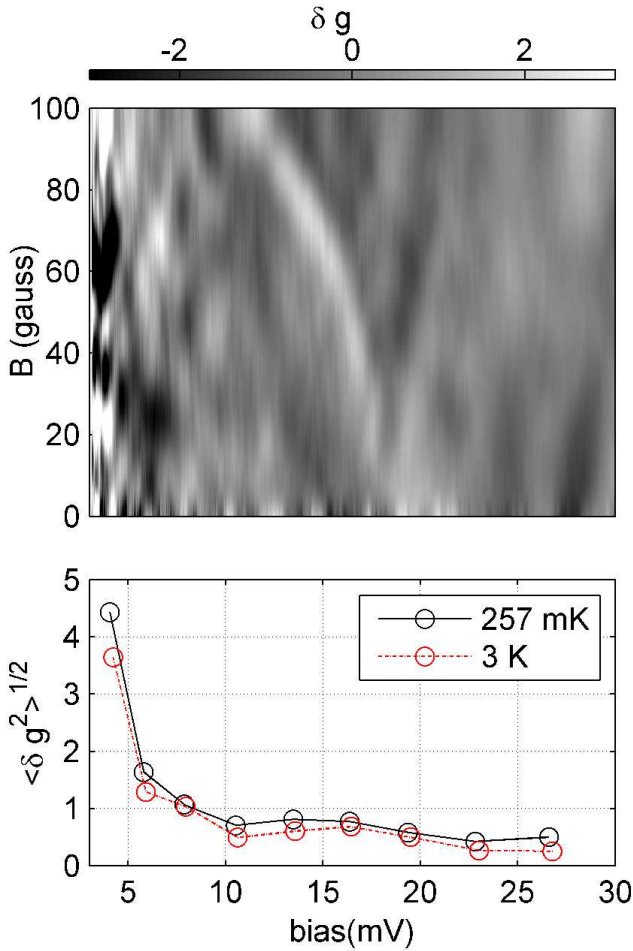}	
\end{center}	
\caption{({\sl color online})Top panel:  grey-color plot of the  fluctuations of the dimensionless conductance  as  a function of the voltage V and of the applied magnetic field H at $T=257mK$. 
Bottom panel: Variance of the dimensionless conductance as a function of the voltage V around zero magnetic field  for T=257mK and T=3K. }
	\label{figpio}
\end{figure}

In Fig.(\ref{figpio})  ({\sl upper panel })  we have reported    the measured  conductance fluctuations  $\: vs \: $  voltage bias and magnetic field at the temperature $T=257mK $, in a grey scale plot. The 
plot shows two different regimes:
\begin{itemize}
\item[a)] 
  Low  voltages ( $V < 3 mV $), where 
fluctuations appear to be very high. Fluctuations  in this range  mostly arise from  precursive switching of the current out of the zero voltage  Josephson state. 
The analysis of this  range of voltages  is better  discussed within the   Macroscopic Quantum Tunneling dynamics\cite{mqt}; it  requires full account of the superconductive correlations and is  not addressed in this paper. 

\item[b)] Large voltages ($V> 5mV $) .
In this regime we observe some  reproducible, non periodic and 
sample dependent fluctuations.
The variance of the conductance $ <(\delta g)^2>_H$,  is plotted   $ vs $ voltage bias ( Fig. (\ref{figpio}) {\sl  bottom panel}) for two temperatures. The scale for its magnitude  is estimated according to  $\delta R/R = \delta g /g \approx  0.01$ with $R= 410 \Omega ( T=300mK) $.
The variance $var[g]$ stabilizes around unity at  $V\gtrsim 7mV$. As the voltage increases   $V \gtrsim 18 $ mV, the variance is increasingly reduced.  
However, small amplitude  fluctuations seem  to persist over a wide  voltage range  up to values which are by far larger than those in normal constrictions. Fluctuations survive up to voltages which are many times the Thouless energy.
\end{itemize}

Here we  focus on the variance $var[g]$  and on the autocorrelation of the magnetoconductance  as a function of  voltage at low temperatures up to  $V >> E_c /e \sim 1 \: mV $. 
Our data can be interpreted on the  basis of models for the  quantum  interference  of carriers   transported  in the narrow diffusive channel  across the GB line.
 The experimental findings are consistent  with a large Thouless energy $E_c$  and quite long dephasing times $\tau _\varphi $. A comparison of the data with the  results of  our  models  seem to confirm  that non equilibrium 
effects induced by the voltage bias  $V$ are not the source of heavy energy relaxation  of the  carriers,
even at voltages $V >> E_c/e $.
A  discussion about the voltage dependence of the variance of the  conductance  for large voltages can be found in the Section VI. 
In the next Section  we analyze  the conductance autocorrelation at finite voltage  in some detail, to extract information about the 
phase coherence length $L_\varphi $ and the phase coherence breaking time $\tau_\varphi$.

\section{Sampling nonlocality: autocorrelation $vs \;\;\Delta H $ }
\begin{figure}[htb]	
\begin{center}		
\includegraphics[width=6cm]{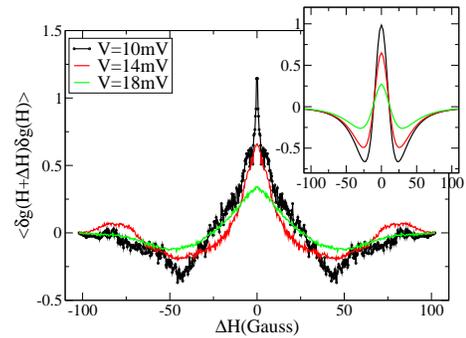}	
\end{center}	
\caption{({\sl color online}) Autocorrelation $<(\delta g )^2>\; vs\; \Delta H $ at 
$T = 257mK $ for three different values of the voltage drop: 
$V=10 \: mV $   {\sl (black curve)}, $V=14 \: mV $ {\sl (red curve)}, $V=18 \: mV $   
{\sl (green curve)}. {\it Inset.} Results of the model calculation sketched in the text for $\xi=1.05$ (black curve),  $1.15$ (red curve), $1.4$ (green curve). }
	\label{fig7}
\end{figure}

The variance $var[g]$,  of the conductance fluctuations
$ \langle (\delta  g )^2 \rangle _H $ discussed in Section III.C  can be derived  from the  maximum at $\Delta H \approx 0 $ of the 
more general autocorrelation   function:
\beq
K_g(V, \Delta H )  
\equiv   \langle \delta  g (V,H+\Delta H ) \delta g(V,H) \rangle _{H}
\label{autoc}
\eneq
The data have been averaged  over H, as usual, as well as over a  
small interval of voltage values  about $V$. 
$K_g$ is the sum of Cooperon  $F_C$ and Diffuson $F_D$ contributions. Due to the independence 
of our results of the temperature we consider the zero temperature limit of Eq. (\ref{var}).
In Fig.(\ref{fig7}) we plot the measured autocorrelation of the dimensionless
conductance vs. $\Delta H $ at  $T = 257mK  $ for   three values of
 $V$: 
 $ 10 \; mV${\sl (blue curve)},  $ 14 \; mV$  {\sl (red curve)} and  
$ 18 \; mV$ {\sl (red curve)} {\sl (color online )}.
The data have been  averaged over a voltage interval of width $\delta V = 0.5mV$ (the results do  not depend  on this choice).

Similar curves have been measured for normal metal wires at zero voltage bias  \cite{oudenaarden}.

The autocorrelation of Fig.(\ref{fig7}) is practically insensitive to increasing  temperature up 
to about $1.5\: K$. 
This fact can be viewed as evidence that a significant contribution to  transport and to the conductance fluctuations is  still provided by the pair current. 
Its time average and absolute value can be seen  as  rather temperature and voltage independent, at fractions of Kelvin, while 
quasiparticles remain rather frozen, provided the voltage does not increase too much\cite{bouchiat}.
For weak links, characterized by and higher barrier transparency the contribution of  the contribution of the supercurrent in the 
I-V curve can be relevant at finite voltages\cite{likharev}. The physical reason is 
that the phase changes in a sharply nonlinear manner with the greater part of the period being close 
to $\pi/2+2 n \pi$. In addition, non-equilibrium effects \cite{lehnert} and unconventional 
order parameter symmetry (with a not negligible second harmonic component in the current-phase Josephson relation
\cite{golubov_RMP}) 
are possible additional sources of supercurrent flowing at finite voltage.
Our results seem to confirm the presence of non negligible contribution of supercurrent at large voltages,
from a different perspective. This is consistent with Ref.s \onlinecite{likharev,lehnert,golubov_RMP} and possibly with the observation of 
fractional Shapiro steps on YBCO grain boundary Josephson junctions\cite{terpstra,early}.
 
According to the remark made above,  we can assume that the current at $T\sim 0$  is only a function  of the phase difference  $\varphi $  between the  two superconducting contacts. This assumes  little  dephasing  induced by  inelastic scattering processes, but not necessarily the absence of
quasiparticle  contributions to the current which still depends on $\varphi$.

The inset of Fig.(\ref{fig7}) shows the result of a simple model calculation of the autocorrelation function based on the following assumptions: $1) $ negligible proximity effect in the sub-micron bridge induced  by the superconducting contacts; $2)$  an  equilibrium 
approach to transport,  in which the current is mostly phase dependent; $ 3) $ handling of the  magnetic field $ H$ is treated  as a small correction and therefore  $H$ only appears in the gauge invariant form of the phase difference.

The model (see  Ref.\cite{teorico} for details) uses as unique fitting parameter  $\xi=  L_y /L_\varphi$, where $L_y $ is the transverse size of the conduction channel.  
The curves plotted in the inset are with $\xi = L_y/L_{\varphi} = $1.05 ({\sl black curve}), 1.15 ({\sl red curve}), 1.4 ({\sl green curve}).
The three different measured curves in the main plot of Fig.(\ref{fig7}) refer to different bias voltages and  cannot be  directly compared to  the theoretical curves  in the inset of Fig.(\ref{fig7}). The qualitative agreement  between experimental and theoretical curves is evident, provided we assume that  $\xi $ increases, with   increasing voltage. This assumption is feasible since, on the one  hand $L_{\varphi}$ is likely  to be reduced when increasing applied voltage, and
the number of conduction channels increases by changing the voltage and,as a consequence, the effective width of the bridge.
If we assume that $L_\varphi$ scales with voltage as $V^{-1/4}$ ( see  discussion in  Section V ),  we find that the $\xi$'s, that have been chosen to draw the inset of Fig.(\ref{fig7}), are consistent with the voltages of the experimental curves within 15\% of error.

The qualitative fit, based on the simple theoretical model used  here,  gives evidence of the fact that non-equilibrium does not seem to spoil the autocorrelation  as a function of the magnetic field, even if the voltage bias exceeds  the Thouless energy: $eV > E_c $. While the dephasing time $\tau _\varphi $  is discussed in the next Section, here we are in position to extract the value of  the phase coherence length $L_\varphi $ from the Power Spectral Density  (PSD) of the conductance autocorrelation function. 

\begin{figure}[htb]	
\begin{center}		
\includegraphics[width=\linewidth]{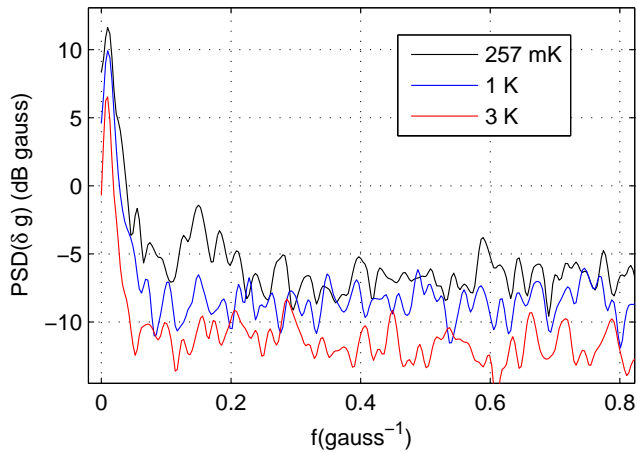}	
\end{center}
	\caption{({\sl color online}) PSD  at  T= 257 mK, 1K, 3K and $V =7 mV$,
averaged over a voltage  interval of $0.5mV$. The curves have been shifted for clarity.}	
	\label{fpsd1}
\end{figure}

The Power spectral density  (PSD) of the conductance autocorrelation  is plotted in Fig.(\ref{fpsd1})  vs $f_H$, the 
conjugate variable to the magnetic field $\Delta H$ , for  data  at T = 273 mK, 1K, 3K and V= 7  mV.
$f_H$ has the dimension of an inverse magnetic field.
Inspection of Fig.(\ref{fpsd1}) shows that there is a linear slope at small frequencies  and a roughly flat trend at larger frequencies. 
The latter is due to the white noise affecting the measurement.  Curves have been shifted, in the figure, for clarity. 
In reality,  both the linear slope and the constant value are almost independent of the temperature.
The linear slope allows to  extract  the value of the phase coherence length $L_\varphi $  according to the fit\cite{hohls}:
\beq
log(PSD/PSD(f_H=0))=-2\pi f_H H_c  + cnst \:\: ,
\label{decibel}
\eneq
as  correlation field $H_c $  can be related to $L_\varphi $ as follows: $L_\varphi \sim \sqrt {h c / 2 e H_c }$. 
From this plot we derive  $H_c \sim 10 \: G $, that  gives  
$ L_\varphi \lesssim 1 \: \mu m $. 
\begin{figure}[htb]	
\begin{center}		
\includegraphics[width=\linewidth]{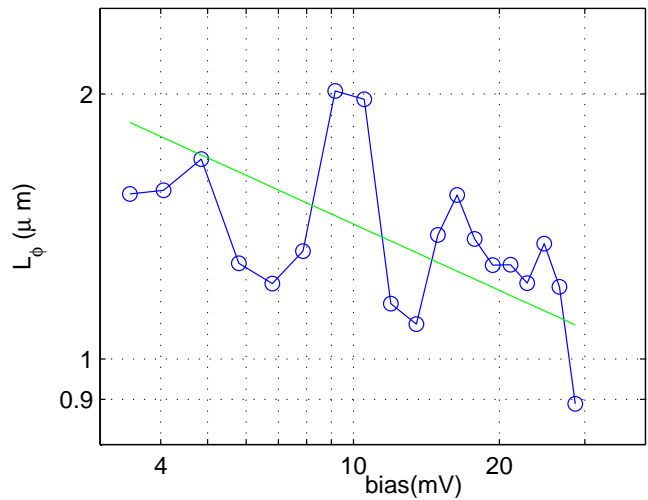}	
\end{center}	
\caption{
Logarithmic plot of $L_\varphi $ vs $V$
derived by using the eq.(\ref{decibel}). Full lines are a guide to the eye. The slope of the straight line is $-1/4$. }
	\label{emilia2}
\end{figure}
A logarithmic plot of the $V$ dependence of  $L_\varphi (V) $  is reported in  fig(\ref{emilia2}). The straight line drawn among the experimental points shows the functional dependence $V^{-1/4}$.

\section{Non equilibrium effects on {\large$\tau_\varphi$}}
\begin{figure}
\centering
\includegraphics[width=\linewidth]{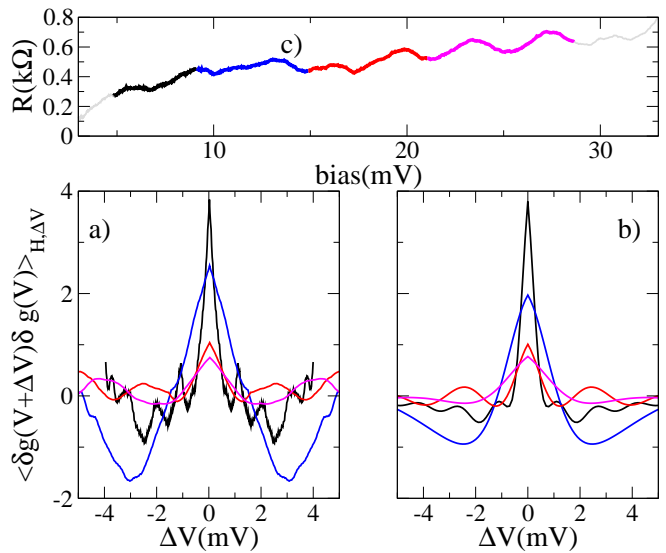}
\caption{  ( {\sl  color online})  The measured  autocorrelation of the conductance 
$K_g(V, H=0,\Delta V )$ 
vs. the voltage difference $ \Delta V $, for   average voltages
$V \approx  7.5,12.5,17.5,25.0 \: mV $ ({ \sl panel } $a)$);  theoretical fit based on Eq.(\ref{sca}) 
  ({\sl panel } $b)$);  Resistance $ R \:  vs. \: V$   ({\sl panel } $c)$). There is  correspondence between the  colored ranges  of   {\sl  panel } $c)$  and the colors of the curves of  the  {\sl  panels } $a)$ and $b)$. }  
\label{autocorri}
\end{figure} 

In this Section we study the  dependence of  the 
autocorrelation   function on $V$ and $\Delta V$:
\beq
  K_g   (V,  \Delta V ) 
\equiv \langle \delta  g (V+\Delta V,H ) \delta g(V,H) \rangle _{H}\:.
\label{expr}
\eneq

A voltage  average has been performed over  intervals centered at $V$, of typical  size lower than the $min\{E_c,\Delta V\}$.

In Fig.(\ref{autocorri}$a$) we report the experimental autocorrelation function $K_g(V, \Delta V )$  vs $\Delta V$ for various values of 
the applied voltage $V$.  We stress that the applied voltage is  larger than  the two natural energy scales: $ E_c /e  \sim 1 \: mV $ and  the nominal superconducting gap $\Delta /e \sim 20 \: mV$. The tail  of the curves  shows a large  anti-correlation dip  when  $\Delta V$ increases and damped  oscillations.
The autocorrelation maximum  flattens  when the applied voltage $V$ increases.
Fig.(\ref{autocorri} $c$) emphasizes  the voltage range to which each curve of the panel $a$ refers, by using  the same  color. 

 We have  reproduced the same trend of the data  in Fig(\ref{autocorri} $a$),by  assuming that  proximity effect induced  by the superconducting contacts does not play  an important role in the transport  and  by adapting the quasi- one-dimensional  non-equilibrium theory of Ref.(\onlinecite{larkin,ludwig}) to our case (see Fig(\ref{autocorri} $b$)). We will  report on  the derivation of our theoretical results  elsewhere\cite{teorico}. We just mention that, our nonequilibrium approach  gives rise to an autocorrelation $K_g (V,\Delta V) $   which is a  function of   $x=e \Delta VL^2/\hbar D$ and of the parameters  $\tau _\varphi /\tau _{C/D} $  and $ (L/L_\varphi )^3 $. Here 
 $ \tau_{C/D}^{-1} =   {e^2L_y^2D(H_1\pm H_2)^2}/({12 \hbar^2 c^2}) $ (with  the  + (-) sign  for the  C (D) case ) is the relaxation time induced by the magnetic field \cite{aleiner}.    
 In  the limit of vanishing   $\Delta V $,  the    result for the variance  is recovered, which,   up to  numerical factors,  is  given in terms of the ratio  between the Airy function $Ai (u) $ and its derivative with respect to the argument $Ai'(u)$: 
    \begin{widetext}
          \beq
         K_g(V,\Delta V =0 ) =  -  {\left (\frac{D\tau _\varphi L_\varphi }{L^3}\right )}\:
         \sum _{\nu = C,D}Ê\: \Re e \left \{  \frac{ Ai 
  \left ( \frac{2\tau _\varphi}{\tau _\nu } 
 \right )} {  \: Ai '  \left ( \frac{2\tau _\varphi}{\tau _\nu}   \right )} \right\}\:\: .
 \label{sca0}
  \eneq    
  \end{widetext}
   $\tif$,$\lvt$ and $\tif/\tau _\nu $ all depend on the voltage.
    This formula is similar to the  thermal equilibrium  result by Altshuler, Aronov, Khmielnitski (AAK)\cite{bla}, which  includes  the  dephasing induced by 
 e-e scattering with small energy transfer. In particular AAK  find 
  \beq
  \frac{1}{\tau _\varphi L_\varphi } =\frac{ e^2 k_B T }{\hbar ^2\sigma }
  \label{iden}
  \eneq
   where $\sigma  $ is the conductivity. At intermediate applied voltages, our  zero temperature prefactor  $D\tif \lvt/L^3 $ in Eq.(\ref{sca0}) looks similar to the AAK prefactor   $\hbar D L_\varphi /(3\pi L^3 k_B T )$ if  $\hbar /\tau _\varphi $ replaces  $k_BT$. 
  
     By choosing  an appropriate  voltage dependence of the  fitting parameter $  {\tau _\varphi} /{\tau _\nu}$ and $\lvt /L $, we find  that the autocorrelation  scales with $\lvt (V) $  as follows:
\beq
     K_g(V,\Delta V ) \approx \sum _{\nu = C,D}æ\:  {\cal{F}}\left [ \frac{\Delta V}{V_c}, 
     \frac{\tau _\varphi} {\tau _\nu},   \left  (\frac{ L_\varphi}{ L}\right )^3  \right ] \:\:\: ,
     \label{sca}
\eneq 
with  $D\tau_\varphi=L_\varphi^2$ and  $\tau _C = \tau _D $, $V_c =\hbar/\tif$. The explicit form of the function ${\cal{F}}$ will be given elsewhere\cite{teorico}. Its limiting form for $ \Delta V \to 0 $ gives Eq.(\ref{sca0}). This scaling law  is exploited to plot the curves of Fig(\ref{autocorri}b).

The scaling among the blue, red and cyan  curves   reproduces reasonably well the experimental pattern.  This indicates that  the  $V$ dependence of  eq.(\ref{sca}) is  well accounted for  by simply reducing  $L_\varphi $ with increasing $V$.
On the contrary, the black curve at the lowest voltage   $V\sim 7.5 mV $ requires adjusting the prefactor  after the scaling,  to make the central peak  higher and narrower.  This could be a hint  to the fact that, when the voltage is rather low, the  superconducting correlations  may   be relevant and should be included in deriving the functional form of the ${\cal{F}}$ function.   
 Needless to say, an increase of the $\tau _\varphi /\tau _\nu $ parameter implies a reduction of the value of  $var[g]$.
 
   In writing  Eq.(\ref{sca}), extra  contributions  arising from the non linear response  have been neglected. Actually,   Eq.(\ref{sca})  would be the result within  linear response theory only,  except for the $V$ dependence 
 of $L_\varphi (V) $. 
   
    The global  interpretation of the data given here shows that a monotonous decrease  of $\tau_\varphi $ with increasing voltage  is not achieved.  A non monotonous decrease  of $var[g]$  vs $V$ is indeed found, as can be seen from Fig.(\ref{figpio}). According to the correspondence $D\tau _\varphi = \lvt ^2 $,  and to Fig.(\ref{emilia2}) a general   decreasing trend of $ K_g(V,\Delta V ) $  with increasing voltage  could take over only above $20 $ mV.    
    In  Eq.(\ref{sca0})  derived from our model calculation,  we have found   $  \hbar /\tif $  in place of  $k_BT $ appearing in  the AAK result of Eq.(\ref{iden}).  
 At larger voltages the expected substitution  in Eq.(\ref{iden}) is \cite{ludwig}   $ k_B  T \to   eV \lvt /L $ and, by  requiring the  consistency, 
 \beq
 \frac{ \lvt}{ L }  \sim \sqrt{\left .  D\tif  \right |_  { eV \lvt /L  } }   \:\:\: \to 
 \left (\frac{\lvt }{L} \right ) ^4 =    \frac{D  }{L^2}\: \frac{\hbar g}{ e V  } . 
   \eneq
   where $ g = \hbar \sigma / (e^2 L ) $ in $1-d $.
 This  would give a  decay law   for the coherence length  $ \lvt \sim V^{-1/4} $, which is not  clearly recognizable in our experiment.

\section{Discussion}

We recollect here the main experimental facts that can be extracted from our data  regarding quantum transport  in  a GB YBCO JJ .

The magnetic dependence of the maximum critical current  suggests an active transport channel of the order of 50-100 nm. Uniformity of the critical current is on scales larger than about 20 nm. The Thouless energy $E_c=\hbar D/L^2$ turns out to be the relevant energy scale in this case.
 The normal resistance $R_N$ of the   HTS junction is of the order of $200 \Omega $, increasing up to $480 \Omega $ with time, due to aging of the sample.  
 The zero field Josephson critical current  appears to  satisfy   $I_C R_N \sim E_c / e $.  This product  is definitely much smaller than $\Delta /e $, where the  nominal superconducting gap is $\Delta =20 \: meV $. This represents   additional  evidence  that  the proximity  effect  induced in  the bridge in the absence of applied voltage is of mesoscopic origin.  The   superconductive pair coherence length  $\xi_s < L_\varphi \lesssim L $,  as opposed to the classical regime, $L <\xi_s $, when  the tail of the order parameter  enters  both  superconductors of the junction\cite{charlat}. 
We speculate that the oscillations in the resistance as  a  function of $V$, shown  in the inset of Fig.(\ref{1res}),  could be due to this mesoscopic  origin. 

Remarkable conductance fluctuations have been found  in a voltage range up to $20 E_c$  in the magnetic  field  range of $ H \in ( -100 , 100) \: G $ for temperatures below  3 K.  In the explored window, we do not measure  an halving of the variance with  increasing  field $H$ \cite{moon}. The crossover field $H_o$  at which the Cooperon contribution to the variance is expected to disappear, as given by Eq.(\ref{vcin})  is  estimated  of the  order of  few Teslas \cite{moon}. Decoherence induced by the Zeeman energy splitting requires even larger fields.  

Transport has been measured in highly non-equilibrium conditions. Hence the  temperature dependence is quite 
weak up to $T \sim 1.5 \: K$.  We have concentrated  our analysis  in the  voltage range   $ V \in (7, 30 )\: mV $, where the  conductance fluctuations reach a steady value for the variance  at  $ T $ below 1 Kelvin,   $var[g](V) \lesssim 1$. These properties confirm that  the fluctuations are due to quantum  coherence at a  mesoscopic  scale.  

The  PSD of the autocorrelation of the  conductance at different fields   $\Delta H$ allows to identify  $H_c \approx 10 \:  G $ as the field  scale  for the mesoscopic correlations, weakly dependent on $V$. This value leads to  a phase coherent  length  $ L_\varphi \lesssim 1\: \mu m$. We have plotted the coherence length extracted from the autocorrelation PSD vs V in Fig.\ref{emilia2} and compared it with  $L_\varphi (V) \sim (V/V_o )^{-s} $ with $ s \sim 0.25$\cite{ludwig}.  A similar exponent has been found in a limited range of voltage bias  $V > E_c/e $  in gold samples \cite{terrier}, in which UCF (universal conductance fluctuations) and Aharonov-Bohm oscillations were found.  Dephasing mechanisms are low frequency electron-electron interaction,  magnetic impurity-mediated  interaction\cite{kaminski} and non equilibrium quasiparticle distribution\cite{pothier}.  According to Fig.\ref{emilia2}, the comparison is not conclusive.
As a matter of fact all data of conductance  autocorrelation at finite voltage reported  for normal wires \cite{oudenaarden,terrier} identify a Thouless energy  $E_c \sim 1\mu eV $, three orders of magnitude smaller than  in our HTS device  and  refer to applied voltages  not larger than $mV$'s.   Still,  mesoscopic  coherence persists  in our sample, up to voltages much larger than the Thouless energy. We do not find any linear increase of the conductance  autocorrelation with voltage at large voltages\cite{larkin}. 

Our model calculation appears to reproduce the gross features in the dependence of the conductance autocorrelation  $ K_g (V,\Delta H,\Delta V) $  on $\Delta H $ as well as on $\Delta V $. In the case of  $ K_g(V,\Delta H,0 ) $ we limit ourselves to the  linear response term only  and the effect of the  voltage bias   just appeared as  a small reduction of $L_\varphi (V) $ with increasing  $V$. 

To  model  the trend  of  $ K_g(V,0,\Delta V ) $  vs $\Delta V$ given by the experiment,   non-equilibrium cannot be ignored. 
Our derivation extends the calculation of  Ref.(\onlinecite{larkin,ludwig}).  We give  a simple estimate of the conductance autocorrelation to fit our experiments.  We invoke the simplest   non-equilibrium distribution for diffusing quasiparticles, that is the collisionless limit\cite{pothier},  by lumping the relaxation processes  in the damping  parameter of the Cooperon/Diffuson propagators $\tau _{C/D}$.  The dependence on the applied voltage is introduced by tuning $\tau_\varphi $. 
We obtain   oscillations in  the  negative tail of the  autocorrelation, $ K_g(V,0,\Delta V ) $ (see Fig.(\ref{autocorri})) and our scaling  procedure fulfills the relation $D\tau _\varphi  = L_\varphi ^2$.
Extra contributions that are specific of the non equilibrium  theory and are known to be responsible first for a  linear increase of the autocorrelation function  with $V$ and subsequently for its power-law decay are not included here.

\section{Conclusions} 
We have reported about  transport measurements  of   high quality biepitaxial Grain Boundary  YBCO  Josephson Junction  at  temperatures below 1 Kelvin, performed over a time period of about 18 months.
A global view on  the data offers a  consistent  picture, pointing to transport across  a single  SNS(superconductor normal superconductor)-like  diffusive  conduction channel  of mesoscopic size $L \lesssim 0.1 \: \mu m $.   We have mostly explored   the magnetoconductance fluctuations in the voltage range  $ eV>>E_c >>k_BT $,where the Thouless energy   $E_c \sim 1 \: meV$.
The Thouless energy, $2-3$ orders of magnitude larger than the one usually experienced in normal mesoscopic or low-$T_c$  superconducting samples, determines  qualitatively the quantum coherent diffusion in the channel. We believe that the oscillations in the resistance that can be seen in Fig.(\ref{1res}) can be due to  quantum  diffusion. 

The mesoscopic correlations are found to be quite robust in our GB narrow channel, even at large voltages. This could not occur if the lifetime of the carriers  were strongly cut by  non-equilibrium relaxation. We conclude that mesoscopic  effects deeply involve superconducting electron-electron correlations, which persist at larger voltages.
Transport features due supercurrents and quasiparticles at finite voltages cannot be disentangled in the pattern of the  conductance, nor in its variance. 
This consideration has led us to approach the problem with  a non-equilibrium model  calculation for generic coherent transport, which highlights the role  of the phase breaking  time $\tau_\varphi $, without including superconducting correlation explicitly.
Fig.\ref{autocorri} shows the comparison between our model results and the autocorrelation experimental data, which is encouraging.
The remarkably long lifetime of the carriers, which we find, appears to be a generic property  in high-$T_c$ YBCO junctions as proved by optical measurements\cite{gedik} and Macroscopic Quantum Tunneling\cite{mqt}.

\begin{acknowledgments}
Enlightening discussions with  I. Aleiner, H. Bouchiat,  V. Falko, A. Golubov,Y. Nazarov, H. Pothier, A. Stern and A. Varlamov at various stages of this work are gratefully  acknowledged.
This work has been partially supported by MIUR PRIN 2006 under
the project \textit{"Macroscopic Quantum Systems - Fundamental
Aspects and Applications of Non-conventional Josephson Structures"},
EC STREP project MIDAS \textit{ "Macroscopic Interference Devices for
Atomic and Solid State Physics: Quantum Control of Supercurrents"} and 
CNR-INFM within ESF Eurocores Programme FoNE -Spintra (Contract No.  ERAS-CT-2003- 980409).
\end{acknowledgments}


\end{document}